\documentstyle[12pt,thmsa,sw20lart]{article}
\input tcilatex
\QQQ{Language}{
American English
}

\begin{document}

\title{Lorentz-Covariant Quantization of Massless Non-Abelian Gauge Fields in The
Hamiltonian Path-Integral Formalism}
\author{Jun-Chen Su \\
%EndAName
Center for Theoretical Physics, Department of Physics,\\
Jilin University, Changchun 130023, \\
People's Republic of China}
\date{}
\maketitle

\begin{abstract}
The Lorentz-covariant quantization performed in the Hamiltonian
path-integral formalism for massless non-Abelian gauge fields has been
achieved. In this quantization, the Lorentz condition, as a constraint, must
be introduced initially and incorporated into the Yang-Mills Lagrangian by
the Lagrange undetermined multiplier method. In this way, it is found that
all Lorentz components of a vector potential have thier corresponding
conjugate canonical variables. This fact allows us to define
Lorentz-invariant poisson brackets and carry out the quantization in a
Lorent-covariant manner.


\end{abstract}

one knows, the massless non-Abelian gauge fields was first quantized by the
elegant Faddeev-Popov approach$^{[1]}$. This approach works in the
Lagrangian (or say, the second order)path-integral formalism. Subsequently,
it was shown that the gauge fields may also be quantized in the perfect
Hamiltonian (or say, the first order)path-integral formalism$^{[2]-[5]}$
along the line of quantization proposed first by Dirac for constrained
systems$^{[6]}$. The Hamiltonian path-integral quantization usually is
performed in the coulomb gauge$^{[2]-[5]}$. Therefore, the quantization and
quantized result are Lorentz-non-covariant. However, the Lorentz-covariant
form of the quantum theory as obtained by the Faddeev-Popov approach is
mostly used in practical applications. Whether and how the massless
non-Abelian gauge field can be Lorentz-covariantly quantized in the
Hamiltonian path-integral formalism? This just is the question we try to
answer in this paper.

Let us recast the Yang-Mills Lagrangian density 
\begin{equation}
{\cal L}=-\frac 14F^{a\mu \nu }F_{\mu \nu }^a
\end{equation}
in the first order form$^{[2]}$. 
\begin{equation}
{\cal L}=E^{ak}\stackrel{\bullet }{A_k^a}+A_0^aC^a-{\cal H}
\end{equation}
where $E^{ak}$ is defined by 
\begin{equation}
E^{ak}=\frac{\partial {\cal L}}{\partial \stackrel{\bullet }{A_k^a}}%
=F^{ako},k=1,2,3.
\end{equation}
which is the field momentum density conjugate to the coordinate $A_k^a$, 
\begin{equation}
C^a=\partial ^kE_k^a+gf^{abc}A^{bk}E_k^c
\end{equation}
and 
\begin{equation}
{\cal H}=\frac 12(E_k^a)^2+\frac 12(B_k^a)^2
\end{equation}
here 
\begin{equation}
B_k^a=-\frac 12\in _{kij}F_{ij}^a
\end{equation}
${\cal H}$ is the Hamiltonian density of the field. In the above, we have
used the Greek letters to denote the four-dimensional indices and the Latin
letters to mark the three-dimensional indices. From the stationary condition
of the action given by the Lagrangian in Eq.(2), it is easy to derive the
following equations of motion$^{[2],[3]}$%
\begin{equation}
\stackrel{}{\stackrel{\bullet }{A_k^a}-\partial
_kA_0^a+gf^{abc}A_0^bA_k^c+E_k^a=0}
\end{equation}
\begin{equation}
\stackrel{}{\stackrel{\bullet }{E_k^a}+\partial
^lF_{kl}^a+gf^{abc}A_0^bE_k^c+gf^{abc}A^{bl}F_{kl}^c=0}
\end{equation}
\begin{equation}
\stackrel{}{C^a\equiv \partial ^kE_k^a+gf^{abc}A_k^bE^{ck}=0}
\end{equation}
The last equation (9) is identified with a constraint condition because
there is no time-derivatives of the field variables in it. This constraint
condition, as shown in the second term in Eq.(2), has already been
incorporated into the Lagrangian by the Lagrange undetermined multiplier
method so as to make the Lagrangian written in Eq.(1) or (2) to be
Lorentz-invariant. The function $A_0^a$ in Eq.(2) acts as a Lagrange
multiplier. Since a massless gauge field has only two polarization states,
among the three pairs of canonical variables $(A_k^a,E_k^a)$ for a given
group index a, only two pairs can be viewed as the independent dynamical
variables. Therefore, in addition to the constraint in Eq.(9), it is
necessary to introduce another constraint so as to eliminate the redundant
degrees of freedom appearing in the theory. There are various choices of the
constraint which are physically equivalent. Commonly, the Coulomb gauge
condition 
\begin{equation}
\partial ^kA_k^a=0
\end{equation}
is preferable to be chosen$^{[2]-[5]}$. The necessity of introducing an
additional constraint implies that the Yang-Mills Lagrangian in Eq.(1)
itself can not give a complete description of the massless gauge field
dynamics unless the constraint in Eq.(10) is combined with it. The
constraint in Eq.(10) may also be incorporated in the Lagrangian by the
Lagrange multiplier method, giving a term $\lambda ^a\partial ^kA_k^a$ in
the Lagrangian. Correspondingly, Eq.(8) will be replaced by 
\begin{equation}
\stackrel{}{\stackrel{\bullet }{E_k^a}+\partial
^lF_{kl}^a+gf^{abc}(A_0^bE_k^c+A^{bl}F_{kl}^c)-\partial _kE_0^a=0}
\end{equation}
where we have set $E_0^a=-\lambda ^a$. Thus, the number of the equations
(7),(9),(10) and (11) is equal to the number of the variables contained in
the equations,  including six canonical variables $(A_k^a,E_k^a)$ and two
Lagrange multipliers $A_0^a$ and $E_0^a$ for a given group index a. This
fact shows self-consistence of the equations.

It is noted that in the Coulomb gauge, the canonical variables can only be
the three-dimensional vectors $A_k^a$ and $E_k^a$ because the variable $%
A_0^a $ has no its conjugate counterpart. In this case, we can only define
the Poisson bracket through the three-dimensional vectors $A_k^a$ and $E_k^a$
and formulate the quantization Lorentz-non-covariantly. In order to perform
the quantization in a Lorentz-covariant manner, instead of the constraint in
Eq.(10), it is suitable to choose the Lorentz gauge condition as the
constraint 
\begin{equation}
\varphi ^a\equiv \partial ^\mu A_\mu ^a=0
\end{equation}
Incorporating this constraint into the Lagrangian shown in Eq.(2) by the
Lagrange multiplier procedure, we have 
\begin{equation}
{\cal L}=E^{a\mu }\stackrel{\bullet }{A_\mu ^a}+A_0^aC^a-E_0^a\varphi ^a-%
{\cal H}
\end{equation}
where 
\begin{equation}
\pi _\mu ^a=\frac{\partial {\cal L}}{\partial \stackrel{\bullet }{A^{a\mu }}}%
=\{ 
\begin{array}{c}
F_{k0}^a=E_k^a,if\text{ }\mu =k=1,2,3; \\ 
-E_0^a,if\text{ }\mu =0.
\end{array}
\end{equation}
are the canonical momentum conjugate to $A_\mu ^a$, 
\begin{equation}
C^a=\partial ^\mu E_\mu ^a+gf^{abc}A^{bk}E_k^c
\end{equation}
and ${\cal H}$ was given in Eq.(5). As we see, in the Lorentz gauge, since
the third term in Eq.(13) contains a time-derivative $\stackrel{\bullet }{%
A_0^a\text{,}}$ the Lagrange multiplier $A_0^a$ has its conjugate variable
provided by the Lagrange multiplier $-E_0^a$. Thus, we have a Lorentz vector 
$E_\mu ^a=(E_0^a,E_k^a)$ and the first term in Eq.(13) and the first term in
Eq.(15) can be written in the Lorentz-invariant form. It is noted that the
terms $E_0^a\stackrel{\bullet }{A}_0^a,$ and $A_0^a\stackrel{\bullet }{E}%
_0^a $ appearing respectively in the first and second terms in Eq.(13) will
be cancelled with each other in the action. Therefore, except for the third
term, the sum of the other three terms in Eq.(13) actually is identical to
the Lagrangian written in Eq.(2). In addition, we note, in order to get
Lorentz-covariant results in later derivations, the second term in Eq.(15)
may also be written in a Lorentz-invariant form 
\begin{equation}
C^a=\partial ^\mu E_\mu ^a+gf^{abc}A^{b\mu }E_\mu ^c
\end{equation}
This is because the added term $gf^{abc}A_0^bE_0^c$ gives a vanishing
contribution to the term $A_0^aC^a$ in Eq.(13) owing to the identity $%
f^{abc}A_0^bA_0^c=0$.

Here we have necessity to discuss the equations of motion derived from the
stationary condition of the action given by the Lagrangian in Eq.(13). These
equations are 
\begin{equation}
\stackrel{}{\stackrel{\bullet }{A_\mu ^a}(x)-\frac{\delta H}{\delta
E_{}^{a\mu }(x)}+\int d^4x\{A_0^b(y)\frac{\delta C^b(y)}{\delta E_{}^{a\mu
}(x)}-E_0^b(y)\frac{\delta \varphi ^b(y)}{\delta E_{}^{a\mu }(x)}\}-\varphi
^a(x)\delta _{\mu 0}=0}
\end{equation}
\begin{equation}
\stackrel{\bullet }{E_\mu ^a}(x)+\frac{\delta H}{\delta A^{a\mu }(x)}-\int
d^4y\{A_0^b(y)\frac{\delta C^b(y)}{\delta A^{a\mu }(x)}-E_0^b(y)\frac{\delta
\varphi ^b(y)}{\delta A^{a\mu }(x)}\}-C^a(x)\delta _{\mu 0}=0
\end{equation}
where the Hamiltonian is defined by $^{[2],[5]}$%
\begin{equation}
H=\int d^4x{\cal H}(x)
\end{equation}
with ${\cal H}(x)$ being given in Eq.(5). In accordance with the definition
of the Hamiltonian and expressions of $\varphi ^a(x)$ and $C^a(x)$ as shown
in Eqs.(12) and (16), the functional derivatives in Eqs.(17) and (18) are
easily calculated. The results are 
\begin{equation}
\frac{\delta \varphi ^a(x)}{\delta A_\mu ^b(y)}=\delta ^{ab}\partial _x^\mu
\delta ^4(x-y)
\end{equation}
\begin{equation}
\frac{\delta \varphi ^a(x)}{\delta E_\mu ^b(y)}=0
\end{equation}
\begin{equation}
\frac{\delta C^a(x)}{\delta A_\mu ^b(y)}=gf^{abc}E^{c\mu }(x)\delta ^4(x-y)
\end{equation}
\begin{equation}
\frac{\delta C^a(x)}{\delta E_\mu ^b(y)}=[\delta ^{ab}\partial _x^\mu
-gf^{abc}A^{c\mu }(x)]\delta ^4(x-y)
\end{equation}
\begin{equation}
\frac{\delta H}{\delta A_\mu ^a(x)}=[\partial
_x^lF_{lk}^a(x)+gf^{abc}A^{bl}(x)F_{lk}^c(x)]\delta _{}^{\mu k}
\end{equation}
\begin{equation}
\frac{\delta H}{\delta E_\mu ^a(x)}=E_k^a(x)\delta _{}^{\mu k}
\end{equation}
By making use of the above derivatives, one may find that Eqs.(17) and (18)
will lead to the equations of motion shown in Eqs.(7) and (11) if we take $%
\mu =k$ in Eqs.(17) and (18) and the constraint equations written in Eq.(12)
and in the following 
\begin{equation}
C^a\equiv \partial ^\mu E_\mu ^a+gf^{abc}A^{b\mu }E_\mu ^c=0
\end{equation}
if we set $\mu =0$ in Eqs.(17) and (18). It should be noted that in the
derivation of the equations of $\mu =0$, we have used the following
equations 
\begin{equation}
\stackrel{\bullet }{A_0^a}(x)-\frac{\delta H}{\delta E_0^a(x)}+\int
d^4y[A_0^b(y)\frac{\delta C^b(y)}{\delta E_0^a(x)}-E_0^b(y)\frac{\delta
\varphi ^b(y)}{\delta E_0^a(x)}]=0
\end{equation}
\begin{equation}
\stackrel{\bullet }{E_0^a}(x)+\frac{\delta H}{\delta A_0^a(x)}-\int
d^4y[A_0^b(y)\frac{\delta C^b(y)}{\delta A_0^a(x)}-E_0^b(y)\frac{\delta
\varphi ^b(y)}{\delta A_0^a(x)}]=0
\end{equation}
which always hold and appear to be identities. It is clear that the
equations given in Eqs.(7),(11),(12) and (26) are sufficient to determine
the eight canonical variables for a given group index which contain four
dynamical variables, two constrained variables and two Lagrange multipliers.
This indicates that the dynamics formulated in the Lorentz gauge is complete.

The canonical structure of the Lagrangian in Eq.(13) implies that the
equations written in Eqs.(12) and (26) which are incorporated into the
Lagrangian by the Lagrange multiplier method can only act as constraint
conditions although there are time-derivatives in them. In contrast to those
constraints given in Eqs.(9) and (10) which are stationary, these
constraints are motional. Let us examine the consistency of these
constraints along the line suggested by Dirac$^{[6]}$. Taking derivatives of
Eqs.(12) and (26) with respect to time and employing Eqs.(17) and (18), we
obtain 
\begin{eqnarray}
&&\{\varphi ^a(x),H\}+\int d^4yA_0^b(y)\{C^b(y),\varphi ^a(x)\}-\int
d^4yE_0^b(y)  \nonumber \\
\times \{\varphi ^b(y),\varphi ^a(x)\} &=&0 \\
&&  \nonumber
\end{eqnarray}
$^{}$%
\begin{eqnarray}
&&\{C^a(x),H\}+\int d^4yA_0^b(y)\{C^b(y),C^a(x)\}-\int d^4yE_0^b(y) 
\nonumber \\
\times \{\varphi ^b(y),C^a(x)\} &=&0 \\
&&  \nonumber
\end{eqnarray}
where the poisson bracket is defined as$^{[5]}$%
\begin{equation}
\{M,N\}=\int d^4x[\frac{\delta M}{\delta A_\mu ^a(x)}\frac{\delta N}{\delta
E^{a\mu }(x)}-\frac{\delta M}{\delta E_\mu ^a(x)}\frac{\delta N}{\delta
A^{a\mu }(x)}]
\end{equation}
Based on this definition and utilizing the derivatives in Eqs.(20)-(25), it
is not difficult to find 
\begin{equation}
\{\varphi ^a(x),\varphi ^b(y)\}=0
\end{equation}
\begin{equation}
\{C^a(x),C^b(y)\}=gf^{abc}C^c(x)\delta ^4(x-y)=0
\end{equation}
where Eq.(26) has been considered, 
\begin{equation}
\{C^a(x),\varphi ^b(y)\}=\partial _x^\mu (D_\mu ^{ab}(x)\delta ^4(x-y))
\end{equation}
where 
\begin{equation}
D_\mu ^{ab}(x)=\delta ^{ab}\partial _\mu ^x-gf^{abc}A_\mu ^c(x)
\end{equation}
\begin{equation}
\{\varphi ^a(x),H\}=-\partial ^kE_k^a
\end{equation}
and 
\begin{equation}
\{C^a(x),H\}=0
\end{equation}
It is emphasized that the nonvanishing of the Poisson bracket in Eq.(34)
implies that the equations (29) and (30) are solvable to the Lagrange
multipliers $A_0^a$ and $E_0^a$. Substitution of the above poisson brackets
into Eqs.(29) and (30) yields 
\begin{equation}
D_\mu ^{ab}\partial ^\mu E_0^b=0
\end{equation}
and 
\begin{equation}
D_\mu ^{ab}\partial ^\mu A_0^b=\partial ^kE_k^a
\end{equation}
These are the second order differential equations for the variables $A_0^a$
and $E_0^a$.

For later purpose, it is useful to consider solutions of the equations (12)
and (26). Noticing the decomposition $A^{a\mu }=A_T^{a\mu }+A_L^{a\mu }$
where $A_T^{a\mu }$ and $A_L^{a\mu }$ are respectively the transverse and
longitudinal components of the vector $A^{a\mu }$ and the transversality
condition $\partial ^\mu A_{T\mu }^a=0$, Eq.(12) may be written as 
\begin{equation}
\partial ^\mu A_{L\mu }^a=0
\end{equation}
Its solution, as is well-known , is 
\begin{equation}
A_{L\mu }^a=0
\end{equation}
Similarly, when the decomposition $E^{a\mu }=E_T^{a\mu }+E_L^{a\mu }$ is
inserted into Eq.(26), noticing the transversality $\partial ^\mu E_{T\mu
}^a=0$ and the solution in Eq.(41), Eq.(26) will be reduced to 
\begin{equation}
\partial ^\mu E_{L\mu }^a+gf^{abc}A_T^{b\mu }(E_{T\mu }^c+E_{L\mu }^c)=0
\end{equation}
Using the expression 
\begin{equation}
E_{L\mu }^a=\partial _\mu Q^a
\end{equation}
where $Q^a$ is a scalar function, we obtain from Eq.(42) 
\begin{equation}
K^{ab}(x)Q^b(x)=\omega ^a(x)
\end{equation}
where 
\begin{equation}
K^{ab}(x)=\delta ^{ab}\Box _x-gf^{abc}A_T^{c\mu }(x)\partial _\mu ^x
\end{equation}
and 
\begin{equation}
\omega ^a(x)=gf^{abc}E_{T\mu }^bA_T^{c\mu }
\end{equation}
With the aid of the Green function(the ghost particle propagator) $%
D^{ab}(x-y)$ which satisfies the equation 
\begin{equation}
K^{ac}(x)D^{cb}(x-y)=\delta ^{ab}\delta ^4(x-y)
\end{equation}
the solution of Eq.(44) is found to be 
\begin{equation}
Q^a(x)=\int d^4yD^{ab}(x-y)\omega ^b(y)
\end{equation}
which is a function of the transverse vectors $A_T^{a\mu }$ and $E_T^{a\mu }$%
. Thus, the function $E_L^{a\mu }$ may be expressed in terms of the $%
A_T^{a\mu }$ and $E_T^{a\mu }$. With the longitudinal components of the
vectors $A_\mu ^a$ and $E_\mu ^a$ being determined by the constraint
equations, the transverse components $A_T^{a\mu }$ and $E_T^{a\mu }$ of the
vectors $A^{a\mu }$ and $E^{a\mu }$ act as the independent canonical
variables. By employing the solutions of the constraint equations,
obviously, the Hamiltonian density denoted in Eq.(5) may be represented via
the independent variables 
\begin{equation}
{\cal H}^{*}(A_{T\mu }^a,E_{T\mu }^a)={\cal H}(A_\mu ^a,E_\mu ^a)_{\mid
\varphi ^a=0,c^a=0}
\end{equation}

Now, we are in a position to formulate the quantization performed in the
Hamiltonian path-integral formalism for the massless non-Abelian gauge
field. According to the general principle of the quantization, we should at
first construct the generating functional of Green's functions via the
independent variables 
\begin{eqnarray}
Z[J] &=&\frac 1N\int D(A_{T\mu }^a,E_{T\mu }^a)\exp \{i\int d^4x[E_T^{a\mu }%
\stackrel{\bullet }{A_{T\mu }^a}  \nonumber \\
&&-{\cal H}^{*}(A_T^{a\mu },E_T^{a\mu })+J_T^{a\mu }A_{T\mu }^a]\}
\end{eqnarray}
In order to express the generating functional in terms of the full vectors $%
A_\mu ^a$ and $E_\mu ^a$, it is necessary to introduce the $\delta $%
-functional $\delta [A_L^{a\mu }]\delta [E_L^{a\mu }-E_L^{a\mu }(A_T^{a\mu
},E_T^{a\mu })]$ into the functional. It is easy to prove that$^{[2],[5]}$%
\begin{equation}
\delta [A_L^{a\mu }]\delta [E_L^{a\mu }-E_L^{a\mu }(A_T^{a\mu },E_T^{a\mu
})]=\det M\delta [\varphi ^a]\delta [C^a]
\end{equation}
where $M$ is the matrix whose elements are 
\begin{equation}
M^{ab}(x,y)=\{C^a(x),\varphi ^b(y)\}
\end{equation}
which were given in Eq.(34). Upon inserting Eq.(51) into Eq.(50) and using
the Fourier representation of the $\delta $-functional 
\begin{equation}
\delta [C^a]=\int D(\eta ^a/2\pi )e^{i\int d^4x\eta ^ac^a}
\end{equation}
We have 
\begin{eqnarray}
Z[J] &=&\frac 1N\int D(A_\mu ^a,E_\mu ^a,\eta ^a)\det M\delta (\partial ^\mu
A_\mu ^a)  \nonumber \\
&&\times \exp \{i\int d^4x[E^{a\mu }\stackrel{\bullet }{A_\mu ^a}+\eta ^aC^a-%
{\cal H}(A_\mu ^a,E_\mu ^a)+J^{a\mu }A_\mu ^a]\}
\end{eqnarray}
Noticing the expression given in Eq.(15), we see, in the above exponent,
there is a $E_0^a$-related term $E_0^a(\partial _0A_0^a-\partial _0\eta ^a)$%
. It allows us to perform the integration over $E_0^a$, giving a $\delta $%
-functional $\delta [\partial _0A_0^a-\partial _0\eta ^a]=\det \left|
\partial _0\right| ^{-1}\delta [A_0^a-\eta ^a]$. The determinant $\det
\left| \partial _0\right| ^{-1}$, as a constant, may be put in the
normalization constant N. The $\delta $-functional $\delta [A_0^a-\eta ^a]$
will disappears when the integration over $\eta ^a$ is carried out. Thus,
considering the expressions given in Eqs.(5),(6) and (15), we can write 
\begin{eqnarray}
Z[J] &=&\frac 1N\int D(A_\mu ^a,E_k^a)\det M\delta [\partial ^\mu A_\mu
^a]\exp \{i\int d^4x  \nonumber \\
&&\times [-\frac 12(E_k^a)^2+E_k^aF_{}^{a0k}-\frac 12F^{akl}F_{kl}^a+J^{a\mu
}A_\mu ^a]\}
\end{eqnarray}
After calculating the Gaussian integral over $E_k^a$, we arrive at 
\begin{eqnarray}
Z[J] &=&\frac 1N\int D(A_\mu ^a)\det M\delta [\partial ^\mu A_\mu ^a]\exp
\{i\int d^4x  \nonumber \\
&&-\frac 14F^{a\mu \nu }F_{\mu \nu }^a+J^{a\mu }A_\mu ^a]\}
\end{eqnarray}
When we employ the familiar expression$^{[1]}$%
\begin{equation}
\det M=\int D(\overline{C}^a,C^a)e^{i\int d^4xd^4y\overline{C}%
^a(x)M^{ab}(x,y)C^b(y)}
\end{equation}
where $\overline{C}^a(x)$ and $C^a(x)$ are the mutually conjugate ghost
field variables and the following limit for the Fresnel functional 
\begin{equation}
\delta [\partial ^\mu A_\mu ^a]=\stackunder{\alpha \rightarrow \infty }{\lim 
}C[\alpha ]e^{-\frac i{2\alpha }\int d^4x(\partial ^\mu A_\mu ^a)^2}
\end{equation}
where $C[\alpha ]=\stackunder{x}{\Pi }(\frac i{2\pi \alpha })^{\frac 12}$
and supplementing the external source terms for the ghost fields, the
generating functional will finally be written in the form 
\begin{eqnarray}
Z[J,\overline{\xi },\xi ] &=&\frac 1N\int D(A_\mu ^a,\overline{C}^aC^a)\exp
\{i\int d^4x[{\cal L}_{eff}  \nonumber \\
&&+J^{a\mu }A_\mu ^a+\overline{\xi }^aC^a+\overline{C}^a\xi ^a]\}
\end{eqnarray}
where 
\begin{equation}
{\cal L}_{eff}=-\frac 14F^{a\mu \nu }F_{\mu \nu }^a-\frac 1{2\alpha
}(\partial ^\mu A_\mu ^a)^2-\partial ^\mu \overline{C}^aD_\mu ^{ab}C^b
\end{equation}
which is the effective Lagrangian for the system under consideration. In
Eq.(59), the limit $\alpha \rightarrow 0$ is implied. Nevertheless, this
limit is unnecessary if we work in general gauges where $\alpha \neq 0$. In
these gauges, the Lorentz condition will be extended to 
\begin{equation}
\partial ^\mu A_\mu ^a-\alpha E_0^a=0
\end{equation}
For this case, it is easy to verify that the poisson bracket $%
\{C^a(x),\varphi ^b(y)\}$ is still given by Eq.(34) and hence the matrix M
remains unchanged. Therefore, when the $\delta $-functional $\delta
[\partial ^\mu A_\mu ^a]$ in Eq.(56) is replaced by $\delta [\partial ^\mu
A_\mu ^a-\alpha E_0^a]$ and then acting on Eq.(56) with the integration
operator $\int D(E_0^a)e^{-\frac i{2\alpha }(E_0^a)^2}$, we still obtain the
generating functional given in Eqs.(59) and (60) with the $\alpha $ being
arbitrary. This generating functional is completely the same as given by the
Faddeev-Popov approach$^{[1]}.$

Up to the present, the Lorentz-covariant quantization in the Hamiltonian
path-integral formalism has been achieved by the novel procedure proposed in
this paper. The essential feature of the procedure is that the Lorentz gauge
condition, as a necessary constraint. is introduced from the beginning and
according to the general procedure established well in mechanics for
constrained systems, it may be incorporated into the Yang-Mills Lagrangian
by the Lagrange multiplier method. In this way, it is found that the
four-dimensional vector potential $A_\mu ^a$ has its four-dimensional
conjugate counterpart. These mutually conjugate canonical variables allow us
to define the poisson bracket and perform the quantization in a
Lorentz-covariant manner. Obviously, the procedure presented in this paper
is more general than the ordinary one because the procedure is suitable to
quantize the gauge field in any gauge. Moreover, by this procedure, one does
not need to make the distinction between the primary constraint and the
second one as well as between the first-class constraint and the
second-class one. The necessary constraints may be chosen from the physical
requirement for the constrained system under consideration.

{\bf Acknowledgment}

The author is grateful to professor Shi-Shu Wu for useful discussions. This
subject was supported in part by National Natural Science Foundation of
China.

{\bf Reference}

[1] L. D. Faddeev and V. N. Popov, Phys. Lett. B25, 29 (1967).

[2] L. D. Faddeev, Theor. Math. Phys. 1, 1 (1970).

[3]E. S. Abers and B. W. Lee, Phys. Rep. C9, 1 (1973).

[4]P. Senjanovic, Ann. Phys. (N. Y.) 100, 227 (1976).

[5]L. D. Faddeev and A. A. Slavnov, Gauge fields: Introduction to Quantum
Theory, The Benjamin Commings Publishing Company Inc. (1980).

[6]P. A. M. Dirac, Lectures on Quantum Mechanics, Belfer Graduate School,
Yeshiva Univ. Press, New York (1964).

\end{document}